# Assessment of different head tilt angles in volumetric modulated arc therapy for hippocampus-avoidance whole-brain radiotherapy

**Cuiyun Yuan[1†], Sisi Xu[1†], Yang Li[1], Enzhuo Quan[1], Dongjie Chen[1], Jun Liang[1*], Chenbin Liu[1*]**

[1]National Cancer Center/National Clinical Research Center for Cancer/Cancer Hospital & Shenzhen Hospital, Chinese Academy of Medical Sciences and Peking Union Medical College, Shenzhen, China

[†]These authors contributed equally to this work and share first authorship

*These authors contributed equally to this work and share corresponding authorship

**\* Correspondence:**
Chenbin Liu, chenbin.liu@gmail.com
Jun Liang, liang23400@163.com

## Abstract

**Purpose:** In the field of radiation therapy for brain metastases, whole-brain hippocampus-avoidance treatment is commonly employed. this study aims to examine the impact of different head tilt angles on the dose distribution in the whole-brain target area and organs at risk. It also aims to determine the head tilt angle to achieve optimal radiation therapy outcomes.

**Methods:** CT images were collected from 8 brain metastases patients at 5 different groups of head tilt angle. The treatment plans were designed using the volumetric modulated arc therapy (VMAT) technique. The 5 groups of tilt angle were as follows: [0°,10°), [10°,20°), [20°,30°), [30°,40°), and [40°,45°]. The analysis involved assessing parameters such as the uniformity index, conformity index, average dose delivered to the target, dose coverage of the target, hot spots within the target area, maximum dose, and average dose received by organs at risk. Additionally, the study evaluated the correlation between hippocampal dose and other factors, and established linear regression models.

**Results:** Significant differences in dosimetric results were observed between the [40°,45°] and [0°,10°) head tilt angles. The [40°,45°] angle showed significant differences compared to the [0°,10°) angle in the average dose in the target area (31.49±0.29 Gy vs. 31.99±0.29 Gy, p=0.016), dose uniformity (1.20±0.03 vs. 1.24±0.03, p=0.016), hotspots in the target area (33.64±0.35 Gy vs. 34.42±0.49 Gy, p=0.016), maximum hippocampal dose (10.73±0.36 Gy vs. 11.66±0.59 Gy, p=0.008), maximum dose in the lens (2.82±1.10 Gy vs. 4.99±0.16 Gy, p=0.016), and average dose in the lens (1.93±0.29 Gy vs. 4.22±0.26 Gy, p=0.008). There is a moderate correlation between the maximum dose in the hippocampi and the PTV length (r=0.49, p=0.001). Likewise, the mean dose in the hippocampi is significantly correlated with the hippocampi length (r=0.34, p=0.04).

**Conclusion:** The VMAT plan with a head tilt angle of [40°,45°] met all dose constraints and demonstrated improved uniformity of the target area while reducing the dose to organs at risk.

Furthermore, the linear regression models suggest that increasing the head tilt angle within the current range of [0°,45°] is likely to lead to a decrease in the average hippocampal dose.

# 1    Introduction

Historically, whole-brain radiation therapy has been the standard treatment for brain metastases to achieve disease control(1, 2). However, the cognitive side effects associated with whole-brain radiation therapy, particularly the detrimental impact on hippocampal function, have raised concerns and highlighted the need for more precise and targeted treatment approaches(3-6). In recent years, whole-brain hippocampus-avoidance (HA-WB) techniques have emerged as promising strategies to minimize radiation-induced damage to the hippocampi while maintaining effective local disease control. These techniques aim to spare the hippocampi, a critical structure involved in memory and cognitive function, from unnecessary radiation exposure(7, 8).

Currently, there are various modalities available for whole-brain hippocampus-avoidance, including intensity-modulated radiation therapy (IMRT)(9), volumetric modulated arc therapy (VMAT)(10, 11), and tomotherapy (TOMO)(12, 13), that have been used in clinical practice to achieve hippocampal sparing during brain metastases treatment. In order to offer further protection to the hippocampi and decrease radiation dose to critical organs, several scholars have introduced a positioning method termed "head tilt," which involves tilting the head at a particular angle(14-19). The findings of the article demonstrate that head tilt can effectively lower the radiation dose to critical organs. Moon et al.(14) and Se et al.(18) compared the dosimetric characteristics between head tilt and no head tilt using the VMAT technique. Their investigation was limited to a single specific tilt angle, and the analysis was restricted to the dose distributions in the target, hippocampi, lens, eyes, and cochlea. Lin et al.(16) employed the couch rotation technique to simulate virtual head tilt angles ranging from 0 to 40 degrees. However, their approach had the potential to introduce errors in organs at risk (OARs) doses due to variations in head tilt angles and modifications in CT cross-sectional scanning. Miura et al.(17) and Chung et al.(15) performed a comparative study using the TOMO technique to assess the quality of treatment plans aiming to spare the hippocampi during whole-brain irradiation at different head tilt angles. However, both studies were limited to investigating specific tilt angles. The aforementioned studies have left the potential effects of a broader range of angles unexplored, and the evaluation of critical organs is also incomplete.

Our study contributed to enhancing the application of HA-WB in a broader range of clinical contexts. To further explore the optimal angle for head tilt, we conducted a comprehensive study to investigate the influence of varying head tilt angles on the radiation dose imparted to the whole brain target volume and critical organs. Additionally, we explored the correlations between the dose spared to the hippocampi and the anatomical geometry. The main objective of this study was to determine the optimal head tilt angle that would lead to optimized clinical outcomes in radiation therapy.

# 2    Materials and methods

## 2.1    Patient selection

In this study, a total of eight patients with brain metastases were included. All patients received whole brain radiotherapy between April 2023 and November 2023 at our institution. All patients had a previous diagnosis of primary tumors originating from the bronchi and lungs. The age range of the study population was 36 to 78 years, with a median age of 57.5 years. The study was approved by the Institutional Review Board (Cancer Hospital Chinese Academy of Medical Sciences, Shenzhen Center Ethics Committee, approval number: SZCHY2023024).

The inclusion criteria for this study are as follows: a) Age ≥ 18 years, no gender restrictions; b) Histologically confirmed primary solid malignant tumor (any type of cancer); c) Karnofsky Performance Score (KPS) ≥ 70; d) Patients who have previously undergone surgical treatment or SBRT/SRS radiotherapy for brain metastases are eligible for inclusion in this study. The following criteria are used to exclude individuals from this study: a) Patients with confirmed leptomeningeal metastases based on imaging; b) Patients who have previously received whole-brain radiotherapy; c) Presence of lesions within a 5mm expansion range of the hippocampus; d) Patients with obstructive hydrocephalus or significant structural deformation of brain tissue confirmed by imaging due to surgery or other benign/malignant brain diseases; e) Patients with severe cervical spondylotic myelopathy or other cervical spine disorders that prevent treatment at the head frame angle; f) Patients with severe underlying diseases (including hypertension, diabetes, heart disease, etc.), or those who are in an acute phase of a certain underlying disease and cannot tolerate radiotherapy.

## 2.2 CT simulation

CT simulation for these patients was performed on a CT scanner (GE Discovery RT; GE Healthcare, Milwaukee, WI). The slice thickness for CT simulation was 2.5mm. Five patients underwent CT scans at tilt angles of 0°, 10°, 20°, 30°, and 40°, while three patients underwent CT scans at tilt angles of 5°, 15°, 25°, 35°, and 45°. All patients were then categorized into five groups based on their tilt angles: [0°, 10°), [10°, 20°), [20°, 30°), [30°, 40°), and [40°, 45°]. Figure 1 illustrates five groups of CT-simulated images from a representative patient. The patients were positioned in the head-first-supine orientation using the Klarity Optek System. They were immobilized in a thermal mask to minimize the inter-fractional and intra-fractional variations. A tilting carbon fiber base plate (Klarity R602-DCF, Guangdong, China) was utilized to elevate the patient's head with a range of 0° to 45°, following the protocol established by our institution.

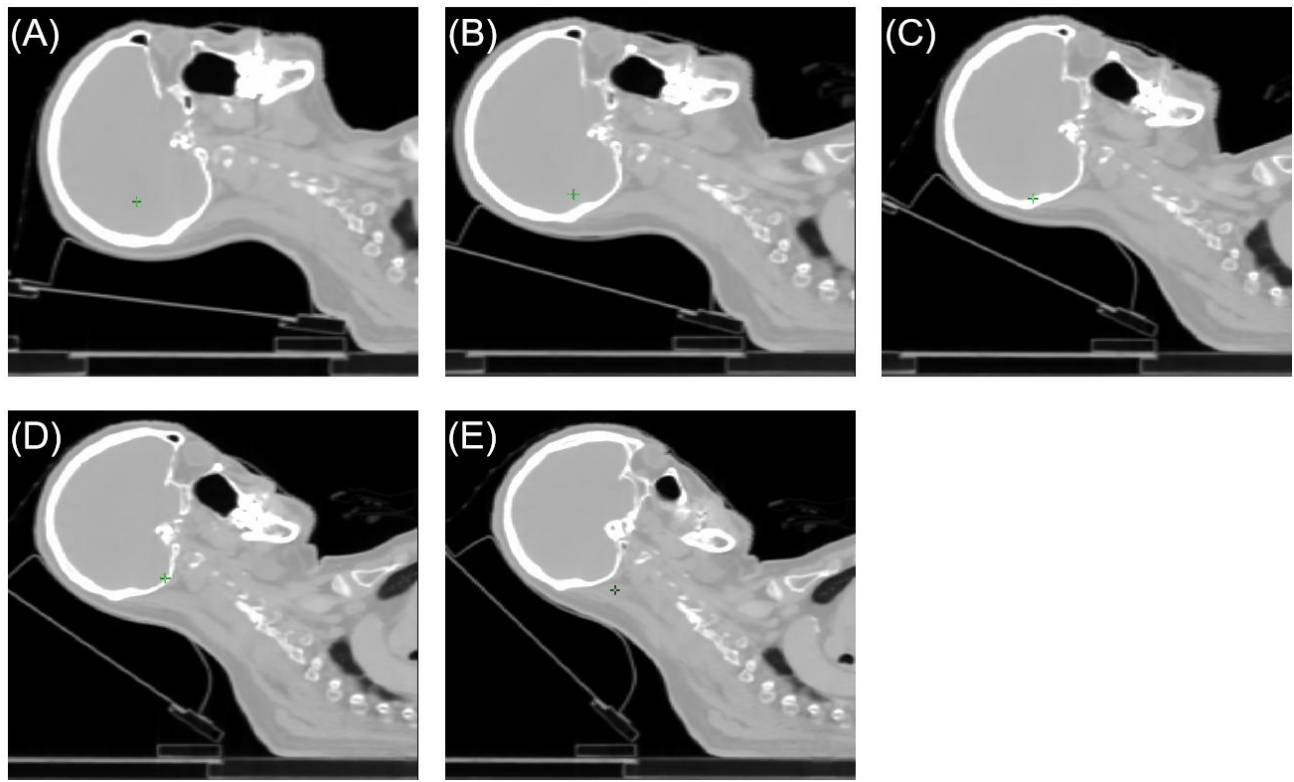

**Figure 1** The sagittal views of a representative patient case. **(A)** [0°,10°) angle group, **(B)** [10°,20°) angle group, **(C)** [20°,30°) angle group, **(D)** [30°,40°) angle group, **(E)** [40°,45°] angle group.

## 2.3 Target and OARs delineation

The target volumes for all groups were demarcated by the physicians. The clinical target volume (CTV) was defined as the whole brain without the hippocampi region, and the planning target volume (PTV) was generated with an extension of 0.3 cm in all three dimensions from the CTV. The surrounding normal OARs, including the hippocampi, lens, eyes, pituitary, brainstem, optic nerve, optic chiasm, and cochlea, were automatically delineated by deep learning contouring software AccuContourTM (Manteia Medical Technologies Co. Ltd., Xiamen, China) and subsequently modified by the physicians. The prescribed dose for the target volume was 30 Gy in 10 fractions according to the Radiation Therapy Oncology Group (RTOG) 0933 protocol(8).

## 2.4 Treatment planning

The treatment plans were generated using the Pinnacle treatment planning system (Philips Medical Systems, Fitchburg, WI, USA). A 6MV FFF energy with a dose rate of 1400MU/min was employed, utilizing the Elekta Infinity linear accelerator (Elekta AB, Stockholm, Sweden) and the volumetric modulated arc therapy (VMAT) technique. The plan consisted of two full arcs, with a collimator angle of 5°. Treatment planning followed the recommendations outlined in RTOG 0933 (Table 1). Maximum and mean dose constraints were applied to limit the dose to the hippocampi, while the dose to the lens was kept below 8Gy. Additionally, efforts were made to minimize the maximum and mean doses to other OARs as much as possible.

**Table 1** Dose constraints for target and organs at risk.

|  |  |
|---|---|
| PTV | $V_{30Gy} \geq 90\%$ |
|  | $D_{2\%} \leq 37.5Gy$ |
| Hippocampi | $D_{100\%} \leq 9Gy$ |
|  | Maximum dose $\leq 16Gy$ |
| Optic nerves and chiasm | Maximum dose $\leq 37.5Gy$ |

## 2.5 Plan evaluation

The dosimetric parameters evaluated for the whole brain target included $D_{2cc}$, $D_{98\%}$, mean dose ($D_{mean}$), homogeneity index (HI), and conformity index (CI)(20). CI represents the degree to which the dose is delivered with high conformity to the target volume while minimizing the dose to surrounding normal tissue and critical organs, and is calculated as follows (Eq.1):

$$CI = \frac{V_{PIL}}{V_{PTV}} \quad (1)$$

Among the given terms, $V_{PIL}$ indicates the volume of the prescription isodose line, and $V_{PTV}$ represents the target volume of the planning target volume (PTV). HI is calculated as the ratio of the dose delivered to 5% of the target volume ($D_{5\%}$) to the dose delivered to 95% of the target volume ($D_{95\%}$)(21). Smaller HI values indicate a more homogeneous irradiation of the planning target volume (PTV). The HI is computed using the following equation: $HI = D_{5\%} / D_{95\%}$. $D_{2cc}$ of the PTV refers to the minimum dose received by at least 2cc of the target volume, while $D_{98\%}$ represents the minimum dose received by at least 98% of the target volume. The dosimetric parameters assessed for OARs included maximum dose ($D_{max}$) and $D_{mean}$. The $D_{max}$ and $D_{mean}$ of the following OARs were evaluated: hippocampi, lens, eyes, optic nerve, pituitary, cochlea, optic chiasm, and brainstem.

## 2.6 Regression model

In order to examine the planning parameters associated with HA-WB treatment plans, a correlation analysis was conducted. We assessed the correlation between hippocampi dosimetric indices ($D_{max}$, $D_{mean}$) and the monitor unit (MU) of treatment plans, PTV length, hippocampi length, and hippocampi angle. Figure 2 illustrates that the PTV length denotes the projection length of the PTV in the sagittal plane, while the hippocampi length signifies the projection length of the hippocampi in the same direction, and the hippocampi angle is the angle between the hippocampi and the horizontal line.

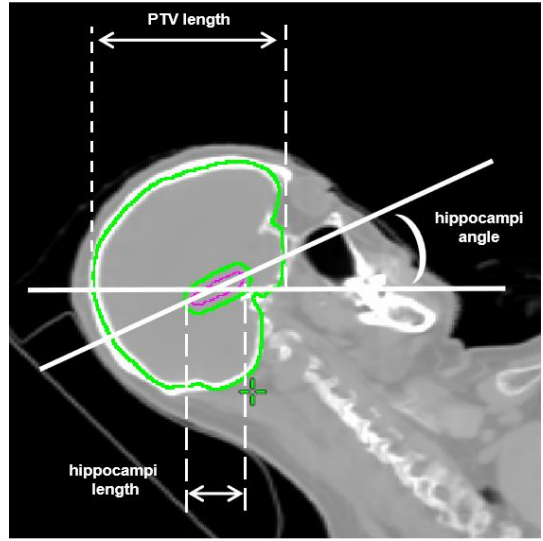

**Figure 2** Illustrations of PTV length, hippocampi length, and hippocampi angle.

Pearson's correlation coefficient (r) was used to calculate the correlation (Eq.2). Linear regression models were constructed using IBM SPSS Statistics software package.

$$\rho_{i,j} = \frac{\sum ij - \frac{\sum i \sum j}{n}}{\sqrt{(\sum i^2 - \frac{(\sum i)^2}{n})(\sum j^2 - \frac{(\sum j)^2}{n})}} \quad (2)$$

in which i represents either the $D_{max}$ or $D_{mean}$ of the hippocampi, j represents one of the following variables: the monitor unit (MU) of treatment plans, PTV length, hippocampi length, or hippocampi angle, n represents the dataset size. The value of ρ ranges from -1 to 1. When ρ=0, i and j are considered uncorrelated. If p=1 or -1, i and j exhibit a linear relationship. The absolute value of ρ determines the degree of correlation, with a higher absolute value indicating a stronger correlation. Conversely, a lower absolute value of the correlation coefficient suggests a weaker correlation.

In this study, the linear regression analysis method(22) was employed to investigate how the dose of the hippocampi can be predicted based on four factors, including MU number of the treatment plan, PTV length, hippocampi length, and hippocampi angle. By conducting regression analysis, the relationship between the dose of the hippocampi and these four factors was elucidated.

## 3 Result

### 3.1 Dosimetric comparison

The summary of dosimetric results for the PTV can be found in Supplementary Table I. Figure 3 presents a comparison of the $D_{mean}$, CI, HI, $D_{2cc}$, and $D_{98\%}$ of PTV across five groups. The $D_{mean}$ of PTV for groups 1 to 5 were reported as follows: 31.99±0.29 Gy, 31.74±0.19 Gy, 31.66±0.24 Gy, 31.73±0.35 Gy, and 31.49±0.29 Gy, respectively. Group 5 demonstrated a significantly higher level of target average dose homogeneity compared to group 1 (p=0.016). In terms of target conformality, the CI values of PTV for groups 1 to 5 were reported as follows: 1.70±0.49, 1.73±0.50, 1.58±0.46, 1.49±0.28, and 1.76±0.42, respectively. The homogeneity index of PTV for groups 1 to 5 were reported as follows: 1.24±0.03, 1.23±0.02, 1.22±0.04, 1.23±0.03, and 1.20±0.03, respectively. Group 5 demonstrated a significantly higher level of homogeneity compared to group 1 (p=0.016), group 2 (p=0.041) and group 4 (p=0.039). The $D_{2cc}$ values of PTV for groups 1 to 5 were reported as follows: 35.78±0.70 Gy, 35.15±0.48 Gy, 35.15±0.59 Gy, 35.04±0.67 Gy, and 34.87±0.50 Gy, respectively. Group 5 showed demonstrated statistically less hot spots compared to group 1 (p=0.016). The $D_{98\%}$ of PTV for groups 1 to 5 were reported as follows: 20.15±0.51 Gy, 19.89±0.29 Gy, 20.23±0.98 Gy, 19.70±0.59 Gy, and 19.91±1.10 Gy, respectively.

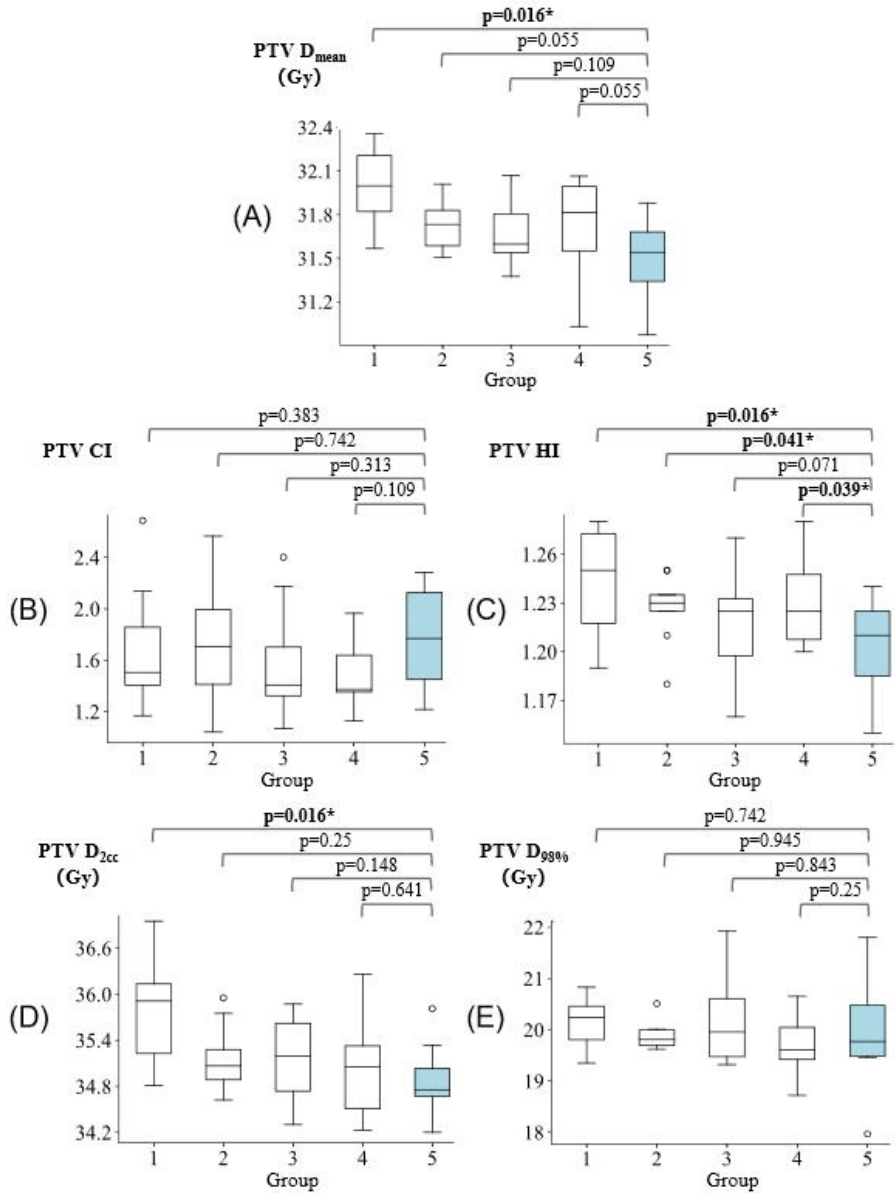

**Figure 3** Comparison of **(A)** PTV mean dose, **(B)** PTV conformity index, **(C)** PTV homogeneity index, **(D)** PTV $D_{2cc}$, and **(E)** PTV $D_{98\%}$ among five groups.

Figure 4 illustrates the $D_{max}$ and $D_{mean}$ values for the hippocampi and lens. The $D_{max}$ of hippocampi for groups 1 to 5 were reported as follows: 11.66±0.59 Gy, 11.40±0.65 Gy, 11.02±0.16 Gy, 11.07±0.19 Gy, and 10.73±0.36 Gy, respectively. Group 5 demonstrated significantly lower values compared to group 1 (p=0.008), group 2 (p=0.016), and group 4 (p=0.039). Regarding the $D_{mean}$ of hippocampi, the values for groups 1 to 5 were: 8.04±0.21 Gy, 7.93±0.16 Gy, 7.99±0.19 Gy, 8.03±0.20 Gy, and 7.97±0.14 Gy, respectively. For the lens, the $D_{max}$ in groups 1 to 5 were reported as 4.99±0.16 Gy, 4.79±0.24 Gy, 4.09±0.59 Gy, 3.25±0.83 Gy, and 2.82±1.10 Gy, respectively. Group 5 showed significantly lower values compared to group 1 (p=0.016), group 2 (p=0.016), and group 3 (p=0.023). The $D_{mean}$ of lens in for groups 1 to 5 were reported as follows: 4.22±0.26 Gy, 3.80±0.30 Gy, 3.06±0.62 Gy, 2.45±0.53 Gy, and 1.93±0.29 Gy, respectively. Group 5 exhibited demonstrated a significantly lower values compared to group 1 (p=0.008), group 2 (p=0.008), group 3 (p=0.008), and group 4 (p=0.039).

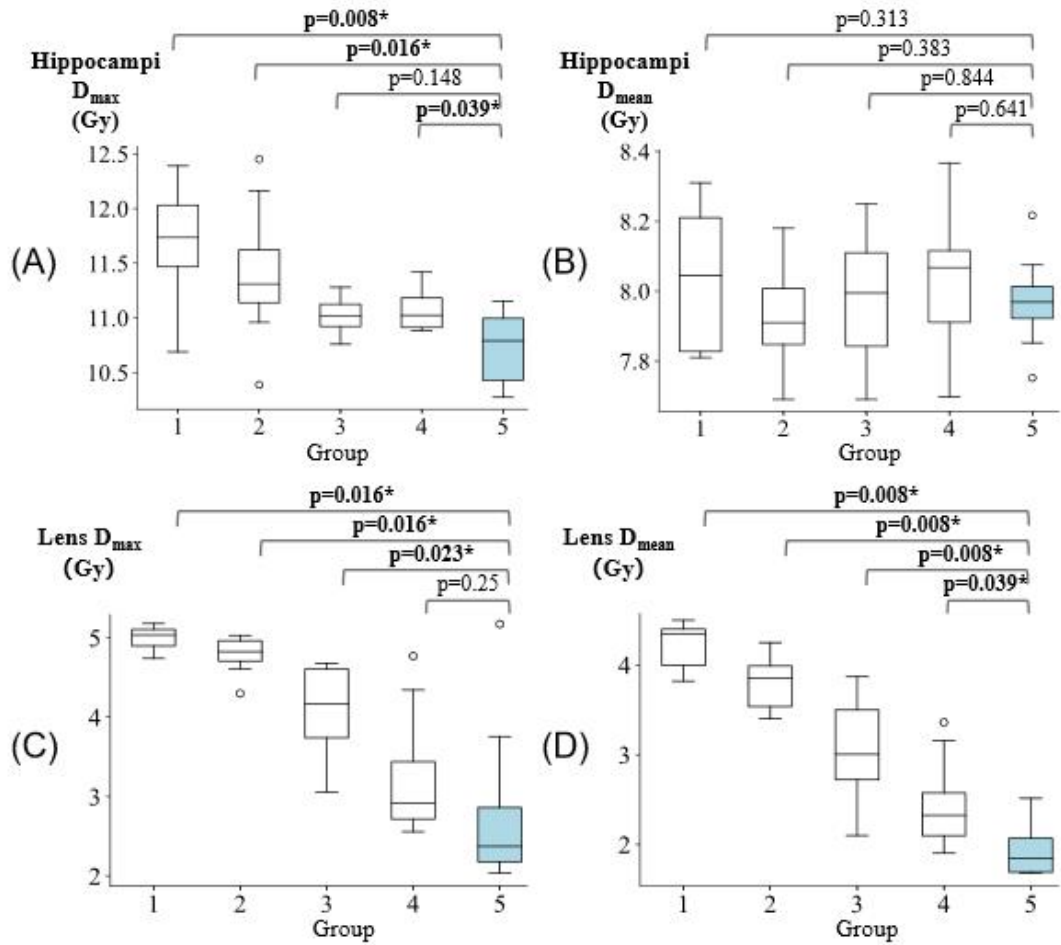

**Figure 4** The comparison of **(A)** hippocampi max dose, **(B)** hippocampi mean dose, **(C)** lens max dose, and **(D)** lens mean dose among five groups.

The dosimetric results for OARs are summarized in supplemental Table II. In Figure 5 and supplemental Table II, The $D_{mean}$ of eyes for groups 1 to 5 were reported as follows: 7.34±0.56 Gy, 7.12±1 Gy, 6.23±0.93 Gy, 5.37±1.07 Gy, and 4.96±0.94 Gy, respectively. Group 5 demonstrated significantly lower values compared to group 1 (p=0.008), group 2 (p=0.016), and group 3 (p=0.039). The $D_{max}$ of optical nerve for groups 1 to 5 were reported as: 34.1±1.17 Gy, 32.6±1.25 Gy, 31.86±0.52 Gy, 31.49±1.66 Gy, and 30.17±2.94 Gy, respectively. Group 5 showed significantly lower values compared to group 1 (p=0.008) and group 2 (p=0.039). The $D_{max}$ of pituitary for groups 1 to 5 were reported as: 34.23±1.69 Gy, 33.12±0.72 Gy, 32.87±0.61 Gy, 32.54±1.31 Gy, and 32.14±2.52 Gy, respectively. Group 5 exhibited significantly lower values compared to group 1 (p=0.039). The $D_{mean}$ of cochlea for groups 1 to 5 were reported as follows: 30.83±0.5 Gy, 30.69±0.65 Gy, 29.73±1.29 Gy, 30.37±1.19 Gy, and 29.83±1.04 Gy, respectively. Group 5 exhibited demonstrated a significantly lower value compared to group 1 (p=0.016). Additionally, no significant differences were observed between Group 5 and the other groups in terms of the $D_{max}$ of cochea, $D_{mean}$ of optical nerve, $D_{max}$ of optic chiasm, $D_{mean}$ of optic chiasm, $D_{max}$ of brainstem, $D_{mean}$ of brainstem, $D_{mean}$ of pituitary, and $D_{max}$ of eyes. Supplementary Table III contains the summary of p-values denoting the significance of differences in PTV and OARs among the five groups.

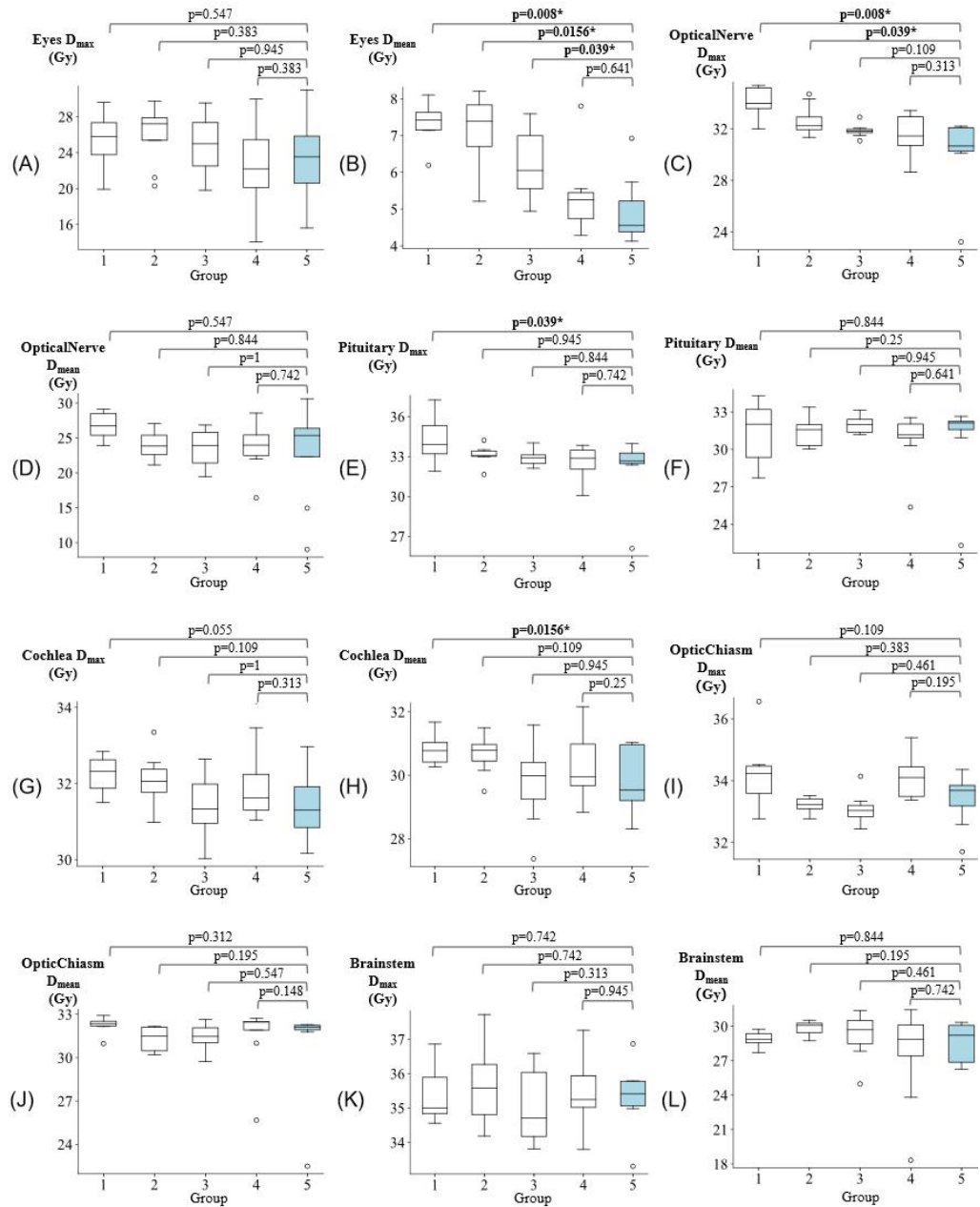

**Figure 5** The comparison of **(A)** eyes max dose, **(B)** eyes mean dose, **(C)** optical nerve max dose, **(D)** optical nerve mean dose, **(E)** pituitary max dose, **(F)** pituitary mean dose, **(G)** cochlea max dose, **(H)** cochlea mean dose, **(I)** optic chiasm max dose, **(J)** optic chiasm mean dose, **(K)** brainstem max dose, and **(L)** brainstem mean dose among five groups.

Figure 6 exhibits the dose distribution for a patient in the sagittal, coronal, and transverse views. The PTV and OARs were assessed across five different groups. The sagittal views from Figure 6 (A) to (E) have tilt angles of 5°, 15°, 25°, 35°, and 45°, respectively. Similarly, the coronal views in Figure 6 (F) to (J) correspond to tilt angles of 5°, 15°, 25°, 35°, and 45°, and the transverse views in Figure 6 (K) to (O) are at the same respective tilt angles. Figure 7 shows the dose volume histogram (DVH) of PTV, hippocampi, optic nerve, optic chiasm, brainstem and lens in five groups. The PTVs have achieved a dose coverage of at least 92% of the prescribed dose. It is apparent that the PTV exhibits similar conformity, and the doses to the hippocampi are also similar. The PTV in the fifth group exhibits better uniformity, while simultaneously having the lowest dose to the lens.

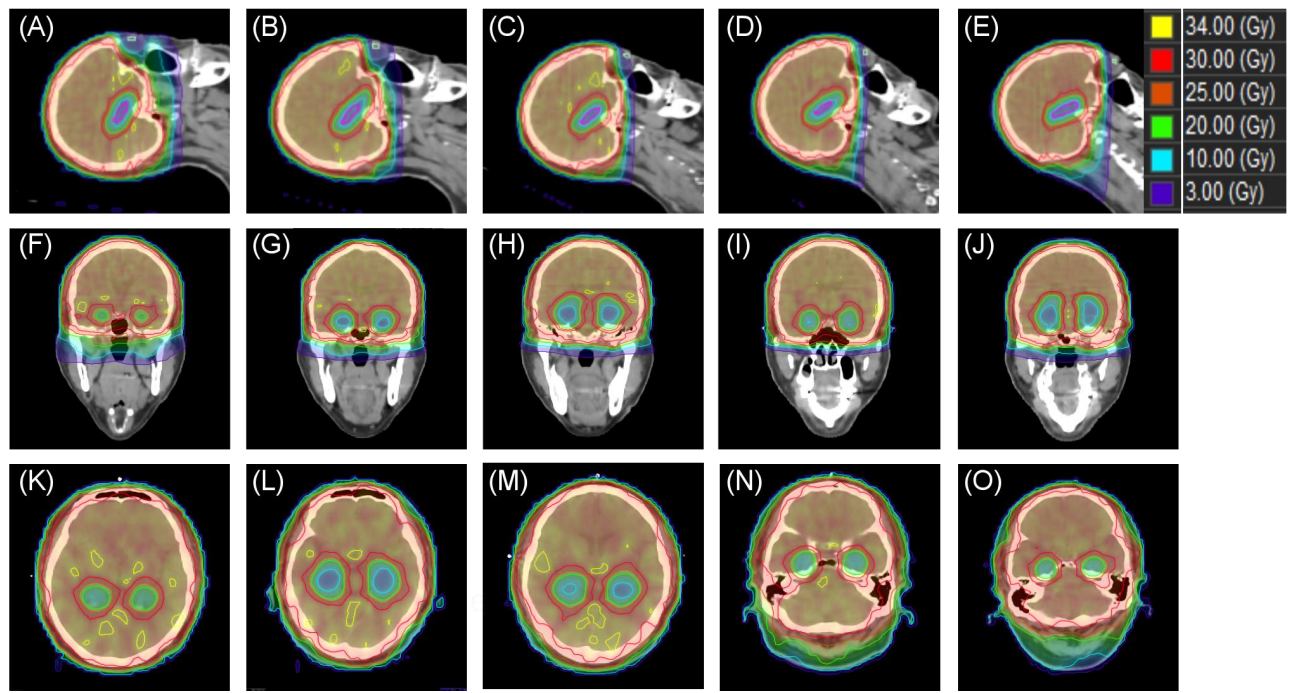

**Figure 6** An example of dose distributions in sagittal, coronal and transverse views with different tile angles. From **(A)** to **(E)** present the sagittal view group 1-5, while **(F)** to **(J)** correspond the coronal view group 1-5 and **(K)** to **(O)** show the transverse view group 1-5.

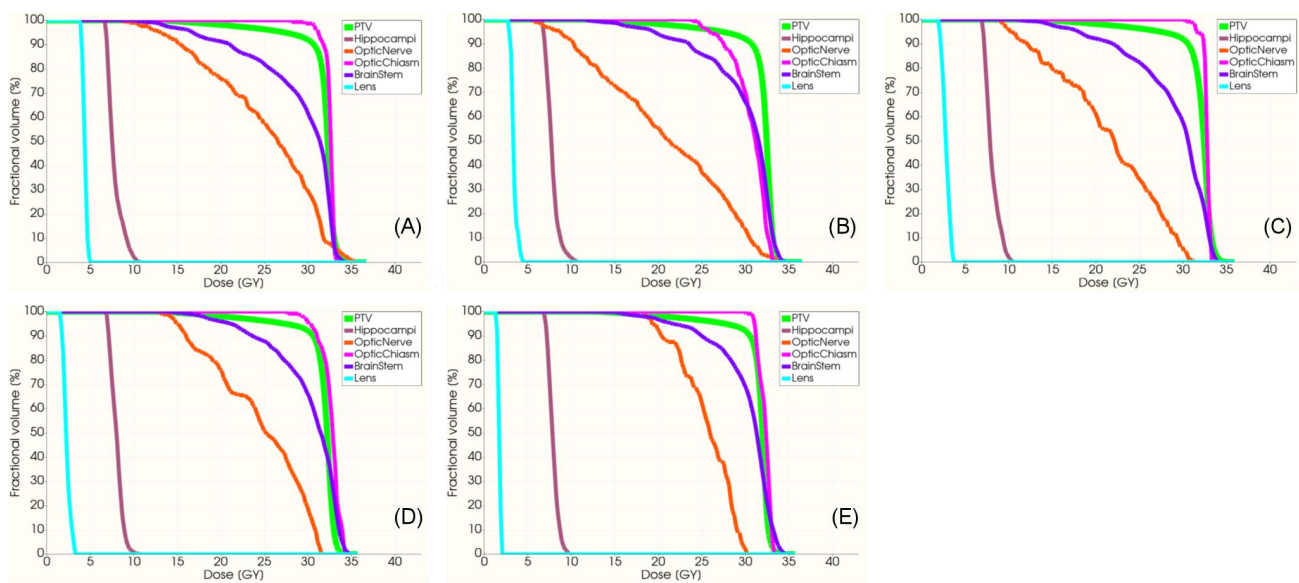

**Figure 7** The DVH of PTV, hippocampi, optic nerve, optic chiasm, brainstem and lens. From **(A)** to **(E)** present the DVH with group 1-5, respectively.

### 3.2 Regression model

We further investigated the correlated factors of the maximum and mean dose in hippocampi. Four treatment planning parameters were considered, including monitor unit (MU) of the treatment plan, PTV length, hippocampi length, and hippocampi angle. Table 2 shows the correlation and significance of the above four factors with hippocampi $D_{max}$. If significance (2-tailed) is less than 0.05, it is considered relevant. The higher the Pearson correlation coefficient (r), the stronger the correlation between this factor and hippocampi $D_{max}$. There is a moderately significant correlation between hippocampi $D_{max}$ and PTV length (r=0.49, p=0.001). Likewise, hippocampi $D_{mean}$ is significantly correlated with the hippocampi length (r=0.34, p=0.04).

**Table 2** The Pearson correlation and significance of hippocampi dosimetric indices were investigated in relation to the following factors: MU, PTV length, hippocampi length, and hippocampi angle.

|  | MU | PTV length | hippocampi length | hippocampi angle |
|---|---|---|---|---|
| hippocampi $D_{max}$ | 0.03 (p=0.88) | **0.49** **(p=0.001)** | -0.19 (p=0.22) | 0.18 (p=0.26) |
| hippocampi $D_{mean}$ | 0.04 (p=0.79) | 0.08 (p=0.62) | **0.34** **(p=0.04)** | -0.26 (p=0.11) |

Based on the aforementioned correlation analysis, we constructed linear regression models to establish the linear formulas for the maximum dose to the hippocampi and PTV length (Eq.3), as well as for the mean dose to the hippocampi and hippocampi length (Eq.4). Supplementary Table IV presents the PTV length, hippocampi length, and hippocampi angle for each patient. The correlation assessment of two linear regression models are shown in Supplementary Table V. The correlation between the maximum dose to the hippocampi and the length of the PTV can be represented by the following Eq.3: the model's R-value is 0.486, and the Durbin-Watson test value is 1.887, indicating a positive correlation between the maximum dose to the hippocampi and the length of the PTV. The ANOVA (Analysis of Variance) yielded an F-value of 11.748, indicating the overall significance of the fitted equation. A higher F-value implies a more substantial equation and a better fit. The significance value is 0.001 (<0.05), affirming the validity of the model.

$$Hippocampi\_D_{max} = 0.374 \times PTV\_length + 5.968 \quad (3)$$

Eq.4 describes the correlation between the mean dose to the hippocampi and its length as follows: the model's R-value is 0.335, and the Durbin-Watson test value is 1.820, implying a positive correlation between the mean dose to the hippocampi and its length. In the ANOVA, the F-value is 4.796, indicating the overall significance of the fitted equation. A higher F-value indicates a more significant equation and a better fit. The significance value is 0.035 (<0.05), signifying the model's validity.

$$Hippocampi\_D_{mean} = 0.111 \times Hippocampi\_length + 7.642 \quad (4)$$

## 4 Discussion

This study provides a thorough dosimetric comparison of the target and various OARs in brain metastasis treatment plans, considering 0°-45° different tilt angles. For the target, within the range of 30-45°, the uniformity of the target dose is satisfactory. Additionally, it should be noted that increasing the head angle may result in a decrease in the dose delivered to both the lens and the hippocampi. Table 3 shows the comparison of head tilt angles, planning techniques, treatment planning systems, and evaluated OARs in our study and previous ones.

Moon et al.(14) and Se et al.(18) conducted a comparative analysis of the dosimetric characteristics between without head tilt and with one tilt angle using VMAT technique. However, the study only investigated a single tilt angle and only analyzed the dose distributions in the target, hippocampi, lens, eyes, and cochlea. Lin et al.(16) employed the couch rotation technique to generate virtual head tilt angles ranging from 0 to 40 degrees. The study demonstrated a similar trend in hippocampal and lens dose compared to our study. However, despite the presence of various angle choices, the virtual angles did not provide information regarding the patient's comfort at each specific angle. Additionally, when compared to the actual head tilt angles, the virtual angles could lead to inaccuracies in OARs as a result of variations in head tilt angles and modifications in CT cross-sectional scanning. Miura et al.(17) and Chung et al.(15) conducted a dosimetric comparison study using TOMO technique to explore the head tilt angles for whole-brain sparing of the hippocampi. However, both studies limited their investigation to specific tilt angles and did not adequately assess the OARs dose sparing. To summarize, our study's strengths include providing a more comprehensive analysis of tilt angles, a more extensive comparison of OARs doses, and establishing a trend for hippocampal dose using a regression model, which could provide guidance for CT simulation and treatment of HA-WB.

**Table 3** Comparison of head tile angles, techniques, planning systems, and OARs

| | head tilt angle | technique | planning system | OARs |
|---|---|---|---|---|
| Ref[Moon] | 30° | IMRT vs VMAT | eclipse | hippocampi, eyes, lens, cochlea |
| Ref[Se] | 40° | VMAT | eclipse | hippocampi, left lens, right lens, both parotid glands |
| Ref[Lin] | 0°-40° | VMAT | eclipse | hippocampi, left lens, right lens |
| Ref[Miura] | 26°-49° | TOMO | TOMO | hippocampi, lens |
| Ref[Chung] | 45° | TOMO | TOMO | hippocampi, eyes, brainstem, optic chiasm, right optic nerves, left optic nerves, normal Brain |
| Our study | 0°-45° | VMAT | Pinnacle | hippocampi, eyes, lens, cochlea, brainstem, optic chiasm, optic nerves, pituitary |

The selection of angle 45 degrees as the maximum head tilt angle in this study stemmed from the fact that the tilting carbon fiber base plate has a maximum tilt capacity of 45 degrees. Additionally, surpassing this angle would have a negative impact on patient comfort. The proposed regression model revealed the relationship between hippocampal dose and the projected lengths of the PTV and hippocampi. The regression model would guide the patient setup in CT simulation and dose constraints in treatment planning. Given the correlation between the anatomical geometric features and hippocampal dose, we would be able to obtain the optimal angle of head tilt in CT simulation and conduct personalized radiotherapy. This approach takes into account the individual patient's anatomy and dosimetric constraints, leading to more accurate and effective radiotherapy planning.

Xue et al.(11) and Yang et al.(23) have demonstrated that by changing the rotation direction of the radiation beam and adjusting the relative position of the hippocampi, non-coplanar plans offer the advantage of reducing the dose to the hippocampi. However, the optimization of non-coplanar plans is more complex due to limitations in gantry angles, increased treatment time, and the risk of potential collisions. Sprowls et al.(24) demonstrated that the use of HyperArc for WBRT with hippocampal sparing offers advantages over coplanar VMAT in terms of hippocampus preservation, reduction in maximum dose, and decreased volume of high dose to the whole brain. Nonetheless, one possible drawback of employing HyperArc is the radiation dose that exits the body when non-coplanar beams pass through the neck and thorax. In contrast, head tilt plans offer the advantage of requiring less time and effort during the planning and treatment stages, being easier to optimize, and effectively decreasing the radiation dose to the hippocampi and neighboring OARs. In head tilt plans, the maximum dose to the lens is 50% lower compared to non-head tilt plans, whereas the decrease in lens dose for non-coplanar plans is comparatively less significant.

Yuen et al.(10) investigated the correlation between collimator selection and the quality of HA-WB plans. However, the study did not examine the impact of collimator selection on the plan. The choice of a 5-degree collimator was based on our institution's expertise in treatment planning of hippocampal sparing whole-brain radiation therapy. Therefore, the potential influence of varying collimator angles on this study was not taken into account. Future research could explore the implications of different collimator angles on the study's conclusions.

The patient's comfort and stability were not evaluated as the head tilt angle increased. However, we believe that a head tilt angle exceeding a certain threshold may affect the convenience of patient positioning and increase the possibility of setup errors during treatment. To ensure proper immobilization throughout the positioning and treatment process, it is recommended to employ a thermoplastic cushion or foam to secure the patient's neck. Figure 8 **(A)-(C)** displays the patient's head tilt positioning for groups 3-5. The red arrows highlight that with an increasing head tilt angle, there is an expansion of the gap between the patient's neck and the immobilization frame, which could lead to discomfort and reduced positioning stability. The selection of head tilt angles in practical clinical settings should prioritize both patient comfort and positioning stability. Moreover, as the head tilt angle increases, the distance between the treatment center and the linear accelerator table also increases, posing a potential risk of collision. When choosing a head tilt angle in clinical settings, the collision risk should be given primary consideration.

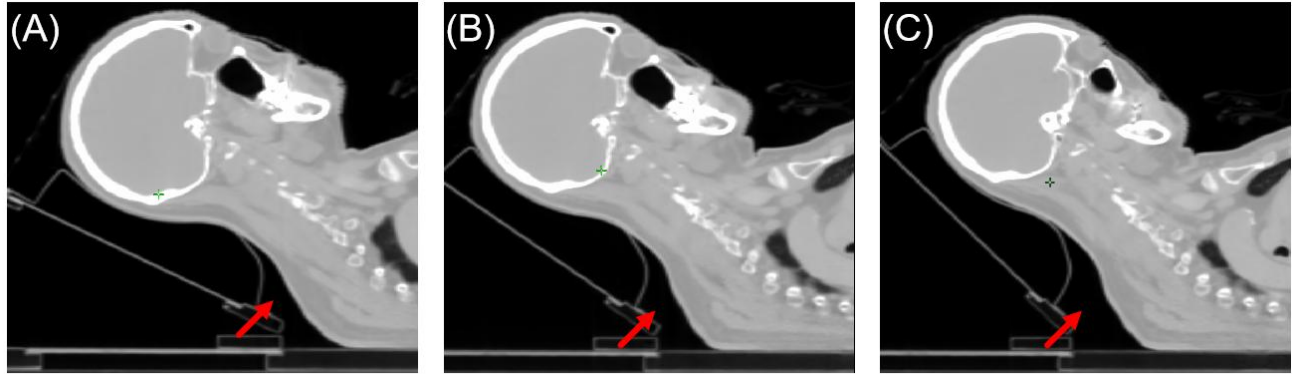

**Figure 8** Illustration of patient positioning comfort. From **(A)** to **(C)** were group 3-5, respectively. The red arrow indicated the level of unsupported neck for the patient.

The relationship between head tilt angle and hippocampal dose in the beam's eye view (BEV) was explored in Figure 9. An increase in head tilt angle is hypothesized to result in a corresponding increase in the hippocampi visible in the BEV, suggesting enhanced modulation of hippocampal dose through MLC optimization at each control point. Furthermore, the adjusted angle leads to a change in the relative position between the lens and the PTV, as depicted in Figure 6, resulting in a substantial decrease in lens dose.

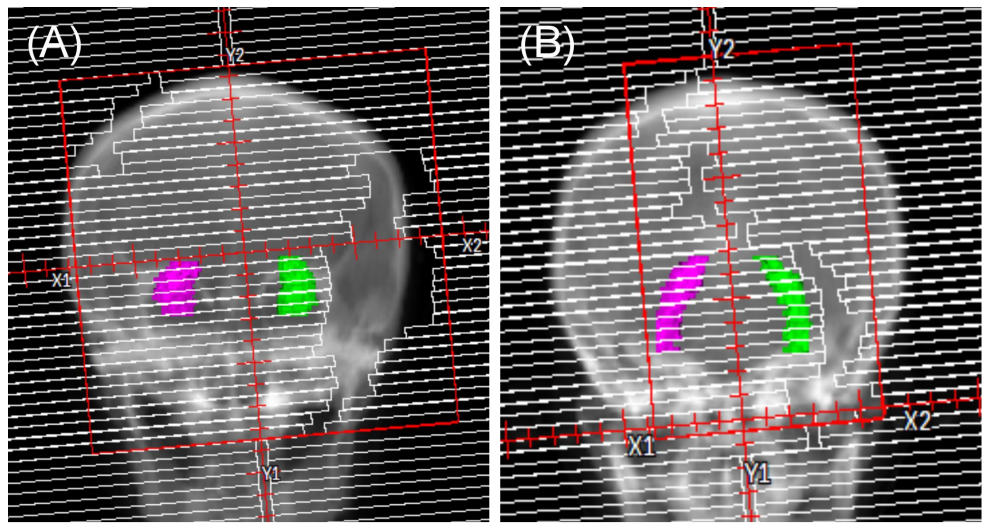

**Figure 9** Illustration of the hippocampi in the beam's eye view. **(A)** 0° head tilt angle and **(B)** 40° head tilt angle.

## 5  Conclusion

The present study demonstrated that VMAT plans with a head tilt angle met all dose constraints. For better dose coverage and uniformity for whole-brain PTV and dose reduction of OARs, the study results suggest utilizing a head tilt angle in [40°,45°]. According to our linear regression models, the hippocampi dose decreases proportionally to the increase in head tilt angle.

## 6 Conflict of Interest

The authors declare that the research was conducted in the absence of any commercial or financial relationships that could be construed as a potential conflict of interest.

## 7 Author Contributions

JL:Conceptualization, Methodology, Investigation, Project administration, Writing-review & editing. CL: Conceptualization, Formal analysis, Investigation, Supervision, Methodology, Writing-review & editing. CY: Data curation, Formal analysis, Methodology, Validation, Software, Writing-original draft. SX: Data curation, Methodology, Investigation, Formal analysis. Writing-original draft. EQ:Conceptualization, Formal analysis, Investigation, Writing-review & editing. YL: Data curation, Formal analysis, Validation, Writing-review & editing. DC:Conceptualization, Formal analysis, Investigation, Writing-review & editing. All authors contributed to the article and approved the final manuscript.

## 8 Funding

This work was supported by National Natural Science Foundation of China (No. 12005301), Guangdong Basic and Applied Basic Research Foundation (No. 2022A1515012456), Natural Science Foundation of Shenzhen City (No: JCYJ20230807150502006), Hospital Research Project (No. SZ2020MS002, E010321015, E010322028), Shenzhen High-level Hospital Construction Fund, Shenzhen Key Medical Discipline Construction Fund (No. SZXK013), Sanming Project of Medicine in Shenzhen (No. SZSM201612063), Shenzhen High-level Hospital Construction Fund., and Shenzhen Clinical Research Center for Cancer (No.[2021]287).

## 9 Data availability statement

The datasets analyzed for this study are not available because of data security requirement of our hospital. Requests to access the datasets should be directed to CL, chenbin.liu@gmail.com

## 10 References

1.	Mehta MP, Khuntia D. Current Strategies in Whole-brain Radiation Therapy for Brain Metastases. Neurosurgery. (2005) 57:S4-33. doi:10.1227/01.Neu.0000182742.40978.E7
2.	Saito EY, Viani GA, Ferrigno R, Nakamura RA, Novaes PE, Pellizzon CA, et al. Whole brain radiation therapy in management of brain metastasis: results and prognostic factors. Radiation Oncology. (2006) 1. doi:10.1186/1748-717x-1-20
3.	Tsai P-F, Yang C-C, Chuang C-C, Huang T-Y, Wu Y-M, Pai P-C, et al. Hippocampal dosimetry correlates with the change in neurocognitive function after hippocampal sparing during whole brain radiotherapy: a prospective study. Radiation Oncology. (2015) 10. doi:10.1186/s13014-015-0562-x
4.	Gondi V, Tomé WA, Mehta MP. Why avoid the hippocampus? A comprehensive review. Radiotherapy and Oncology. (2010) 97:370-6. doi:10.1016/j.radonc.2010.09.013
5.	H.Suh J. Hippocampal-Avoidance Whole-Brain Radiation Therapy: A New Standard for Patients With Brain Metastases? Journal of Clinical Oncology. (2014) 32:3789-91. doi:10.1200/JCO.2014.58.4367
6.	Brown PD, Gondi V, Pugh S, Tome WA, Wefel JS, Armstrong TS, et al. Hippocampal Avoidance During Whole-Brain Radiotherapy Plus Memantine for Patients With Brain Metastases:

Phase III Trial NRG Oncology CC001. Journal of Clinical Oncology. (2020) 38:1019-29. doi:10.1200/jco.19.02767

7. Stoker J, Vora S, Patel A, Grosshans D, Brown PD, Vern-Gross T, et al. Advantages of intensity modulated proton therapy during hippocampal avoidance whole brain radiation therapy. Physics and Imaging in Radiation Oncology. (2018) 8:28-32. doi:10.1016/j.phro.2018.11.001

8. Gondi V. Preservation of memory with conformal avoidance of the hippocampal neural stem-cell compartment during whole-brain radiotherapy for brain metastases (RTOG 0933): a phase II multi-institutional trial. Journal of Clinical Oncology. (2014) 32:3810.

9. Pokhrel D, Sood S, Lominska C, Kumar P, Badkul R, Jiang H, et al. Potential for reduced radiation-induced toxicity using intensity-modulated arc therapy for whole-brain radiotherapy with hippocampal sparing. Journal of Applied Clinical Medical Physics. (2015) 16:131-41. doi:10.1120/jacmp.v16i5.5587

10. Yuen AHL, Wu PM, Li AKL, Mak PCY. Volumetric modulated arc therapy (VMAT) for hippocampal-avoidance whole brain radiation therapy: planning comparison with Dual-arc and Split-arc partial-field techniques. Radiation Oncology. (2020) 15. doi:10.1186/s13014-020-01488-5

11. Xue J, Jin S, Zhang H, Zou K, Sheng J, Tang J, et al. A simplified non-coplanar volumetric modulated arc therapy for the whole brain radiotherapy with hippocampus avoidance. Frontiers in Oncology. (2023) 13. doi:10.3389/fonc.2023.1143564

12. Shimizu H, Sasaki K, Aoyama T, Tachibana H, Koide Y, Iwata T, et al. Parotid gland dose reduction in the hippocampus avoidance whole-brain radiotherapy using helical tomotherapy. Journal of Radiation Research. (2022) 63:55-62. doi:10.1093/jrr/rrab107

13. Takaoka T, Tomita N, Mizuno T, Hashimoto S, Tsuchiya T, Tomida M, et al. Dosimetric Comparison of Helical Tomotherapy and Intensity-Modulated Proton Therapy in Hippocampus- and Scalp-Sparing Whole Brain Radiotherapy. Technology in Cancer Research & Treatment. (2021) 20. doi:10.1177/15330338211060170

14. Moon SY, Yoon M, Chung M, Chung WK, Kim DW. Comparison of the extent of hippocampal sparing according to the tilt of a patient's head during WBRT using linear accelerator-based IMRT and VMAT. Physica Medica. (2016) 32:657-63. doi:10.1016/j.ejmp.2016.04.009

15. Chung Y, Yoon HI, Ha JS, Kim S, Lee IJ. A Feasibility Study of a Tilted Head Position in Helical Tomotherapy for Fractionated Stereotactic Radiotherapy of Intracranial Malignancies. Technology in Cancer Research & Treatment. (2016). doi:10.7785/tcrt.2012.500420

16. Lin YF, Chen PH, Shueng PW, Lin HH, Lai LH. Evaluation of various head flexion angles in hippocampal-avoidance whole-brain radiotherapy using volumetric modulated arc therapy. Radiation Physics and Chemistry. (2020) 173. doi:10.1016/j.radphyschem.2020.108884

17. Miura K, Kurosaki H, Utsumi N, Sakurai H. Use of a Head-Tilting Baseplate During Tomotherapy to Shorten the Irradiation Time and Protect the Hippocampus and Lens in Hippocampal Sparing-Whole Brain Radiotherapy. Technology in Cancer Research & Treatment. (2021) 20. doi:10.1177/1533033820986824

18. Se An Oh, Ji Woon Yea JWP, Jaehyeon Park. Use of a head-tilting baseplate during volumetric-modulated arc therapy (VMAT) to better protect organs at risk in hippocampal sparing whole brain radiotherapy (HS-WBRT). Plos One. (2020) 15. doi:10.1371/journal.pone.0232430

19. Shimizu H, Sasaki K, Aoyama T, Tachibana H, Tanaka H, Koide Y, et al. Examination of the best head tilt angle to reduce the parotid gland dose maintaining a safe level of lens dose in whole-brain radiotherapy using the four-field box technique. Journal of Applied Clinical Medical Physics. (2021) 22:49-57. doi:10.1002/acm2.13151

20. E S. Radiation Therapy Oncology Group: radiosurgery quality assurance guidelines. International Journal of Radiation Oncology* Biology* Physics. (1993) 27:1231-9.
16

21. Weiss E, Siebers JV, Keall PJ. An analysis of 6-MV versus 18-MV photon energy plans for intensity-modulated radiation therapy (IMRT) of lung cancer. Radiotherapy and Oncology. (2007) 82:55-62. doi:10.1016/j.radonc.2006.10.021
22. Su X, Yan X, Tsai CL. Linear regression. WIREs Computational Statistics. (2012) 4:275-94. doi:10.1002/wics.1198
23. Yang B, Liang Y, He S, Liu Y, Zhang K, Qiu J. Dosimetric comparison of coplanar and noncoplanar volumetric modulated arc therapy for hippocampal-sparing whole-brain radiation therapy. Medical Dosimetry. (2023) 49:85-92. doi:10.1016/j.meddos.2023.08.010.
24. Sprowls CJ, Shah AP, Kelly P, Burch DR, Mathews RS, Swanick CW, et al. Whole brain radiotherapy with hippocampal sparing using Varian HyperArc. Medical Dosimetry. (2021) 46:264-8.